\documentclass[sigconf]{acmart}

\AtBeginDocument{%
  \providecommand\BibTeX{{%
    \normalfont B\kern-0.5em{\scshape i\kern-0.25em b}\kern-0.8em\TeX}}}
\usepackage{tabularx}
\usepackage{bbding}
\usepackage{graphicx}
\usepackage{geometry}
\usepackage{subfigure}
\usepackage{amsmath}
\usepackage{makecell}
\usepackage{float}
\usepackage{color}
\usepackage{booktabs}
\usepackage{caption}
\usepackage{hyperref}

\copyrightyear{2024}
\acmYear{2024}
\setcopyright{acmlicensed}\acmConference[CHI '24]{Proceedings of the CHI
Conference on Human Factors in Computing Systems}{May 11--16,
2024}{Honolulu, HI, USA}
\acmBooktitle{Proceedings of the CHI Conference on Human Factors in
Computing Systems (CHI '24), May 11--16, 2024, Honolulu, HI, USA}
\acmDOI{10.1145/3613904.3642133}
\acmISBN{979-8-4007-0330-0/24/05}

\author{Yuanning Han}
\authornote{These authors contributed equally to this work.}
\email{12010843@mail.sustech.edu.cn}
\orcid{0009-0002-4978-8532}
\affiliation{
\institution{Southern University of Science and Technology}
\city{Shenzhen}
\country{China}}

\author{Ziyi Qiu}
\authornotemark[1]
\email{213192675@seu.edu.cn}
\orcid{0009-0008-6341-0218}
\affiliation{
\institution{Southeast University}
\city{Nanjing}
\country{China}}

\author{Jiale Cheng}
\email{19085542d@connect.polyu.hk}
\orcid{0009-0005-7275-0809}
\affiliation{
\institution{Hong Kong Polytechnic University}
\city{Hong Kong}
\country{China}}

\author{RAY LC}
\authornote{Correspondences should be addressed to LC@raylc.org.}
\email{LC@raylc.org}
\orcid{0000-0001-7310-8790}
\affiliation{
\institution{City University of Hong Kong}
\city{Hong Kong}
\country{China}}

\begin{document}
\begin{sloppypar}

\title[When Teams Embrace AI]{When Teams Embrace AI: Human Collaboration Strategies in Generative Prompting in a Creative Design Task}

\begin{abstract}
Studies of Generative AI (GenAI)-assisted creative workflows have focused on individuals overcoming challenges of prompting to produce what they envisioned. When designers work in teams, how do collaboration and prompting influence each other, and how do users perceive generative AI and their collaborators during the co-prompting process? We engaged students with design or performance backgrounds, and little exposure to GenAI, to work in pairs with GenAI to create stage designs based on a creative theme. We found two patterns of collaborative prompting focused on generating story descriptions first, or visual imagery first. GenAI tools helped participants build consensus in the task, and allowed for discussion of the prompting strategies. Participants perceived GenAI as efficient tools rather than true collaborators, suggesting that human partners reduced the reliance on their use. This work highlights the importance of human-human collaboration when working with GenAI tools, suggesting systems that take advantage of shared human expertise in the prompting process.

\end{abstract}




\captionsetup{font=footnotesize, labelfont=bf, textfont=normalfont}
\begin{teaserfigure}
    \centering
    \includegraphics[width=1\linewidth]{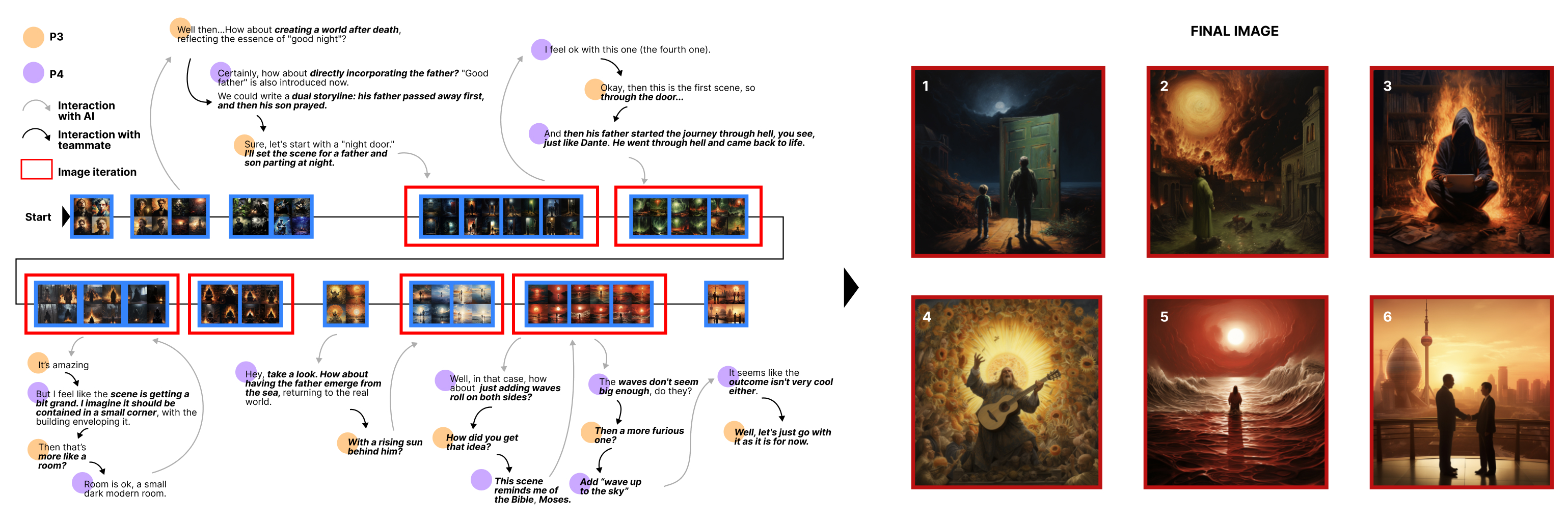}
    \caption{The process (left) and outcomes (right) of two human participants working as a team to use GenAI to iteratively draft the stage design for a performance based on the theme of a Dylan Thomas poem.}
    \Description{This figure shows the whole workshop process of P3 and P4: with the participation of GenAI (Midjourney), they eventually generated six sketches of stage design based on a given poem through multiple communications and attempts.}
    \label{teaser}
\end{teaserfigure}

\maketitle

\section{Introduction}\label{sec:Introduction}
Generative AI (GenAI) has been employed in creative tasks wherein humans collaborate with AI in endeavors such as creative writing \cite{co-writingScreenplays, Metaphoria19Gero}, drawing \cite{ItIsYourTurn, WhenIsATool}, performance \cite{EmbodyingTheAlgorithm}, game design \cite{FriendCollaborator}, art exhibition \cite{lc2023human,lc2023together}, and music arrangement \cite{AIMusic20Louie}. AI-generated content can facilitate the human creative process by providing inspiration, offering novel ideas, facilitating human expression, and performing laborious tasks \cite{ILead,FriendCollaborator, Metaphoria19Gero}. However, how does GenAI work within teams of human collaborators? Existing research primarily focuses on individual prompting with GenAI in creative tasks, which is a challenging endeavor, especially for non-expert single users \cite{userprompting}. Yet, the prompting process is vital to yield desirable outputs in GenAI. How do diverse human teams work together to overcome these challenges when interacting with the prompting process in creative collaborative tasks?


In addition to providing us with generative content for creative purpose, research shows that GenAI tools have the potential to facilitate decision-making \cite{decisionmaking, AIdecisionmaking} and consensus-building \cite{codesign} when they are involved in a collaboration scenario. One study found that compared with human teams that did not enhance their coordination over time, teams collaborating with AI were able to improve steadily \cite{teamconflict}. Research into the detailed dynamics of team-AI collaboration is needed to understand how best to adapt to diverse voices in collaborative processes. This allows us to develop strategies for optimizing decision-making, contribution, and efficient cooperation in teams that increasingly work with GenAI.

Whether we can achieve the best performance when collaborating with AI depends on multiple factors, such as trust or acceptance towards AI \cite{AItrust}. For example, we can significantly develop users' trust and increase the acceptance of AI by increasing its interpretability or explainability \cite{Trustworthy, Explainability}, thereby enhancing human-AI collaboration. Studying a team's perception of GenAI performance can lead to designs for more harmonious and efficient interactions, for example conversational interventions or system mechanics for increasing the perception of reliability or transparency.

Team collaboration is often required in multidisciplinary creative tasks such as the example of art design for stage performance \cite{stagedesign, lc2023contradiction, lc2023active}.
Stage design involves artists, directors, designers, scriptwriters, and performers working in a cross-disciplinary environment where an artist talks with the performer, or performer with a designer, etc. The outcomes in stage design have both visual and textual elements, with visuals often developed from texts like a film from a screenplay. Art design for stage performance provides a case study of collaborative practice where collective ideation is necessary. Inspired by both its collaborative nature and the potential of utilizing GenAI during the creative process, we chose stage design as the task for our study.

We define co-prompting in this study (collaborative prompting) as a process that involves sharing and discussing prompts for GenAI systems among two or more individuals. We adopted a qualitative approach to investigating co-prompting in creative tasks:

\vspace{0.2cm}
\textbf{RQ1:} \textit{What are the challenges teams face when co-prompting generative AI tools during creative design tasks, and what strategies do they employ to overcome them?}

\textbf{RQ2:} \textit{How do individuals perceive the roles of GenAI and human collaborators in co-prompting generative AI tools for creative design?} 
\vspace{0.05cm}

We conducted an online workshop where participants teamed up in pairs to design artwork for a stage performance. The design process involved co-prompting Midjourney based on inspiration from the poem "Do Not Go Gentle Into That Good Night," with an option of using ChatGPT (GPT-3.5). We used semi-structured interviews to probe each participant after the workshop and performed thematic analysis on the initial coding of this qualitative information to obtain insights into the co-prompting process. 

The results suggest that co-prompting enables participants to overcome challenges associated with building and adjusting prompt words. The prompting process remains challenging, and communication costs within the team may increase. Nevertheless, co-prompting fosters a setting where participants feel encouraged to experiment and share their ideas, facilitating more in-depth discussions regarding creative content and enhancing mutual understanding. However, discussions on prompting strategies could also take participants away from meaningful conversations about creative ideas. Participants noted a shift in their attention towards guiding GenAI and selecting its output in an attempt to alleviating laborious production. They also expected GenAI to provide more unforeseen inspirations rather than desirable content that matches their imagination. During the creative process, participants actively sought their collaborators' opinions on the prompting process and the generated output to evaluate GenAI ability.

We then discuss the challenges and strategies associated with using GenAI in collaborative scenarios. We propose that the co-prompting process could be a double-edged sword, as it may facilitate creative ideation on the one hand to support team collaboration, but also add mental demands of strategic prompting, which could reduce team performance. Athough users would sometimes trust and rely on GenAI to complete tasks, they still appreciate and prioritize ideas of their own or those of their human teammates.

This study identifies strategies of humans working with other humans in the co-prompting creative process, and explores how collaboration and prompting could mutually foster each other. We offer design insights for developing collaborative creative design systems that cater to the needs of multiple users and suggest leveraging human-human collaboration in the GenAI creative process.

\section{Background}\label{sec:Background}
\subsection{Human-AI Creative Collaboration}
Recent work in HCI have applied GenAI tools into the creative process. These studies suggest that AI has the potential to provide users with unexpected ideas \cite{ItIsYourTurn}. The tasks include comprehension and creative writing \cite{co-writingScreenplays, Writing1, Writing2, yang2022ai}, 
mixed-initiative storytelling \cite{sun2022bringing},
artistic visual creation \cite{drawing1, ILead, Drawing2, Drawing3}, audio production \cite{Music1, Music2, Music3}, video generation \cite{Video}, etc. One study explored involving more than one GenAI agent in using ChatGPT to generate a story and visualize it with Stable Diffusion \cite{gpt-ImagePrompt2}.


The prompting process introduces challenges for non-expert users who often struggle with how to get started, how to choose the right instructions, or how to build up and modify the prompts \cite{userprompting}. Previous research has revealed that artists desire additional assistance and guidance when constructing prompt words and refining details \cite{promptartist}. While visual representations are especially needed in creative design \cite{DesignSketches}, designers find it challenging to utilize text-to-image GenAI to translate one's original ideas into precise prompt words while ensuring comprehensiveness and accuracy, often resulting in images of diverse quality \cite{CollaborativeDiffusion}. 

Many strategies have been applied for the practice of improving of prompt engineering. Prompt tools like "Promptify", which utilize a suggestion engine powered by large language models, can help users quickly explore and craft prompts \cite{Promtify}. Another tool is designed to assist with the prompting process through automatic prompt editing \cite{promptediting}.
In addition to these tools, prompting strategies for common users such as "try multiple generations to get a representative idea of what prompts return" and "focus on keywords rather than the phrasings of the prompt" have also proven useful \cite{Designguidance}. Seeking help on prompting from the web has also been observed as a common strategy for users \cite{userprompting}.


In summary, previous work have shown that users encounter challenges during the prompting process, which may be overcome through tools for prompting or useful prompting strategies. However, most of these findings are based on a single person using GenAI. Thus, we intend to investigate what challenges, strategies, and dynamics develop when team-AI collaboration in creative processes involve the interaction of other humans.

\subsection{Perception of AI in Creative Processes}

Previous work have investigated how GenAI is perceived during the creative process. Participants who often engaged in co-creation with GenAI described AI agents as their friends or co-workers instead of mere tools, as if GenAI is incorporated into creative teams \cite{WhenIsATool, SandInTheLoop}. Some view AI as a creative partner, finding it to be a source of inspiration that enhances their overall work experience \cite{FriendCollaborator}. Humans do not to appear to be able to distinguish between AI and human-generated text, indicating that GenAI can effectively reproduce the nuance of the text written by humans \cite{lc2021imitations}.

Nevertheless, there are concerns about the risks and limitations related to GenAI. There have been raised concerns about bias, stereotype, plagiarism, diminished creative ideas, and worries about being replaced by GenAI tools \cite{co-writingScreenplays}. Researchers have also indicate that current tools have their limitations and can only act as information providers instead of decision-makers \cite{lai2023exploring}. The focus in process-oriented human-AI interaction research appears to be on facilitating collaboration between humans and AI, rather than on using GenAI as a replacement for human creativity \cite{CollaborativeDiffusion}.

Existing studies of AI perception predominantly stem from individual interactions with GenAI. In regards to the dynamics of team-AI collaboration, other questions arise: How will people perceive AI when another human is in the team? Will there be a difference in collaborating with humans and AI during the creative process?

\subsection{Human Collaboration in Stage Design}

To investigate the dynamics of team-AI collaboration, interactions between human team members themselves must be considered. An example of humans-only collaboration in interdisciplinary design is art design for the stage. Factors like cognitive diversity, shared mental model, and open communication influence the experience of designing for the stage and its final performance.

Effective teamwork is critical for the stage design process. Stage designers are typically members of a design team, which may include the director, lighting designer, costume designer, sound designer, stage manager, music director, choreographer, and playwright or librettist \cite{stagedesign}. The interdisciplinary nature of the design team may stimulates ideation by enhancing cognitive diversity \cite{cog1, cog2}. Stage designers' work begins with a careful reading of the script and identification of all the prop elements needed for the story. This is followed by building rough models for discussion in meetings with the core design team \cite{stagedeisgn2}. During the meetings, designers with diverse experiences need to reach a consensus on the general setting of the show, with sketches and drawings of the stage created to illustrate the concepts \cite{stagedeisgn2}. The creation of a shared mental model within a team to build common output from individual ideas has been identified as an effective method for enhancing team performance and satisfaction \cite{share1, share2, share4}. Open communication in this process can also enhance creative output \cite{communi1, communi2}. However, challenges also emerge in collaboration, most notably how designers integrate their individual collaborative methods into a design team, and how the design team as a whole works to translate the play from written words into visual art on stage \cite{stagedesign3}.

Reflecting on the potential of GenAI in interdisciplinary team-AI creative collaboration, we chose the stage design process from script reading to conceptual sketching as the creative task to study. The focus of the work is on team-AI collaboration, rather than promoting applications of GenAI in the stage design industry.

\section{Methods}\label{sec:Methods}

\subsection{Participants and Recruitment}
We posted a call for participants on social media platforms and also sent it via researchers' personal communication channels. All participants did not have experience extensive knowledge in working with GenAI. We recruited 18 subjects (8 males, 10 females), all either holding or pursuing an undergraduate degree, with 12 out of 18 majoring in art and design (industrial design, product design, architecture design, painting), 3 of 18 having a background in performance (dancing), and one expressing interest in pursuing stage design professionally. All participants had design or performance background and adequate command of English and were of Chinese ethnicity. Participants were randomly assigned into 9 pairs. They gave their consent to participate and to have their data collected anonymously. All study procedures conformed to the institutional IRB guidelines on human subject study, and the data collected was analyzed while being blind to the subjects' identities.

The participants we recruited were predominantly students with experience in design, art, or performance, though none had a professional background in stage art design. This decision was guided by specific considerations. Our study's primary objective was not to examine the application of GenAI in stage art design; instead, we aimed to investigate the effects of GenAI on team collaboration in creative tasks. We selected stage art design as our study's focus because it epitomizes creative collaborative work. We believe that our findings have broader implications and are applicable to various forms of creative collaboration.

\begin{figure}[ht]
    \centering
    \includegraphics[width=1\linewidth]{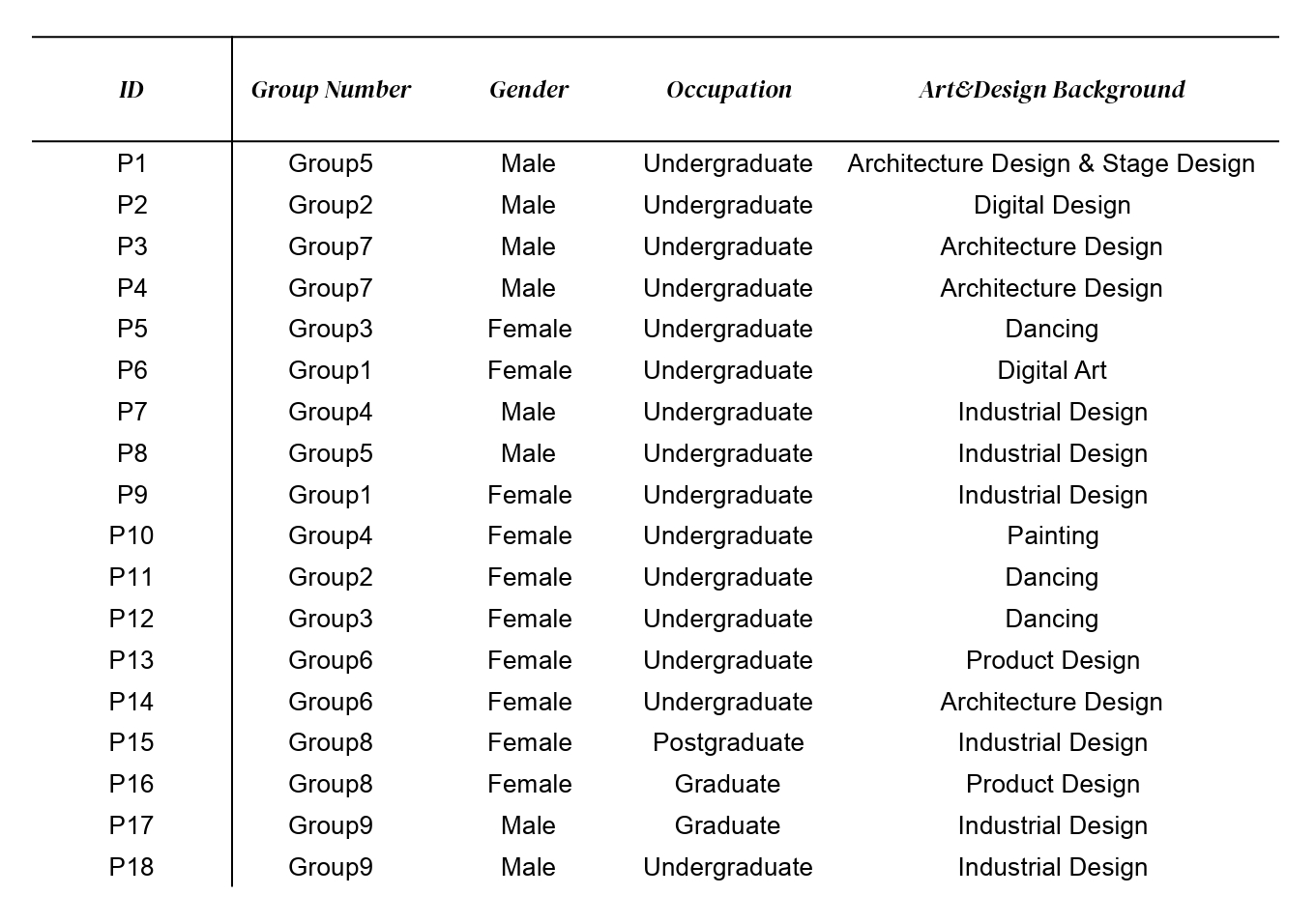}
    \caption{Demographic information of the participants, including group number, gender, occupation, and their art or design backgrounds. The 18 participants were divided into 9 groups, with a total of 10 females and 8 males. All participants were either current students or recent graduates.}
    \label{fig:enter-label}
    \Description{This table shows the demographic information of the participants, including their ID, group number, gender, occupation, and whether or not they had an art or design background. The 18 participants were divided into 9 groups, with a total of 10 females and 8 males. All participants were either current students or recent graduates, and all had an art or design background.}
\end{figure}

\subsection{Workshop Design}
\begin{figure}[ht]
    \centering
    \includegraphics[width=1\linewidth]{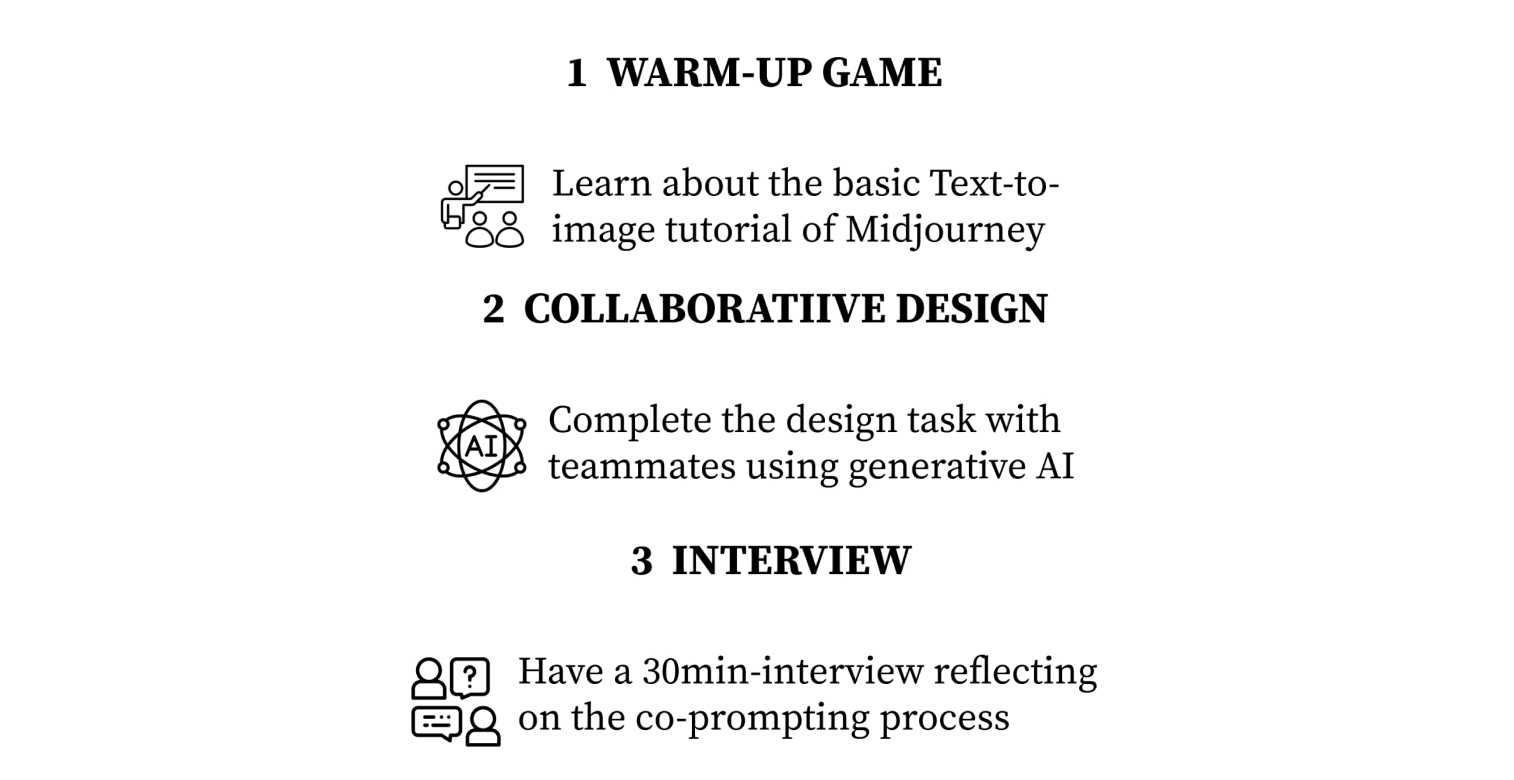}
    \caption{The three sessions experienced in the design process: (1) A warm-up game to learn about test-to-image GenAI, (2) Collaborative design with other teammates using GenAI together, (3) Posthoc interview to gather insights regarding the task.}
    \Description{This Figure shows the flow of the workshop, which had 3 phases. The first stage was a warm-up game, mainly to familiarize the participants with the basic functions of Midjourney; the second stage was the main design collaboration stage, where two participants would use ChatGPT and Midjourney to complete the corresponding design tasks, and the researcher was not involved in the whole.}
    \label{fig:Workshop process}
\end{figure}

The online workshop (2 hours) asked participants to create five or more stage design sketches or references based on inspiration drawn from the Dylan Thomas poem "Do Not Go Gentle Into That Good Night." The workshop took place on the Tencent Meeting platform. To generate the sketches, participants had access to both ChatGPT (GPT-3.5) and Midjourney during the process (Figure \ref{fig:Workshop process}).

Each workshop included a warm-up session to familiarize participants with text prompting, which is the only function that they used when generating images on Midjourney during the workshop. They could generate as many images as they wanted but were asked to select five works as their design output. Screen sharing was needed when using GenAI, and the whole process was screen-recorded. Access to Midjourney and ChatGPT presented challenges in mainland China, as not all participants can access these platforms on their computers. If a participant was unable to access these platforms, they were offered remote control access through Tencent Meeting. On the other hand, if access was available, one participant in each group would use their computer to access Midjourney and ChatGPT, sharing their screen during the meeting. The participant tasked with operating ChatGPT and Midjourney was encouraged to actively listen to their collaborators' ideas during the prompting process. Due to potential network issues, participants may request to discontinue remote control access and instead ask the researcher to operate Midjourney and ChatGPT temporarily. In such cases, participants communicated their prompts to the researcher until the network problem was solved. Researchers observed the entire work process of the workshop, only intervening when necessary (e.g., technical assistance or answering questions). Participants were encouraged to think aloud and share ideas with teammates during the workshop.

Considering the interdisciplinary nature of the design process and the necessity of collaboration amongst individuals from different backgrounds, we chose stage design as the task as it often takes inspiration from other art forms. We gave the entire text of the poem to participants beforehand, and asked them to transform the inspirations from the poem into a stage design with sketches or references. We used Midjourney for text-to-image after pilot testing with other tools, due to its easy-to-use interface and ease of understanding for non-expert users. The way people work with AI in design is already demanding, so we aimed to minimize the additional cognitive load during the process.

\subsection{Interview Protocol}
The semi-structured interviews were conducted remotely via Tencent Meeting with each workshop participant individually at the conclusion of each intervention. Each interview lasted between 20 and 30 minutes. Before the start of the interview, all participants were informed that their discussions would be recorded and transcribed. During the interview session, the researcher asked participants to recall and evaluate their experiences using GenAI in teamwork. They were prompted to identify challenges they faced in collaboration, describe their strategies to overcome these challenges, and express their feelings when encountering such issues. Additionally, participants were encouraged to assess the contributions made by GenAI and their human collaborators to their teamwork and to reflect on their own performance during the workshop. All interviews were conducted in Mandarin, subsequently recorded and transcribed using Tencent Meeting, and translated by the researchers into English.

\begin{figure*}[ht]
    \includegraphics[width=0.7\linewidth]{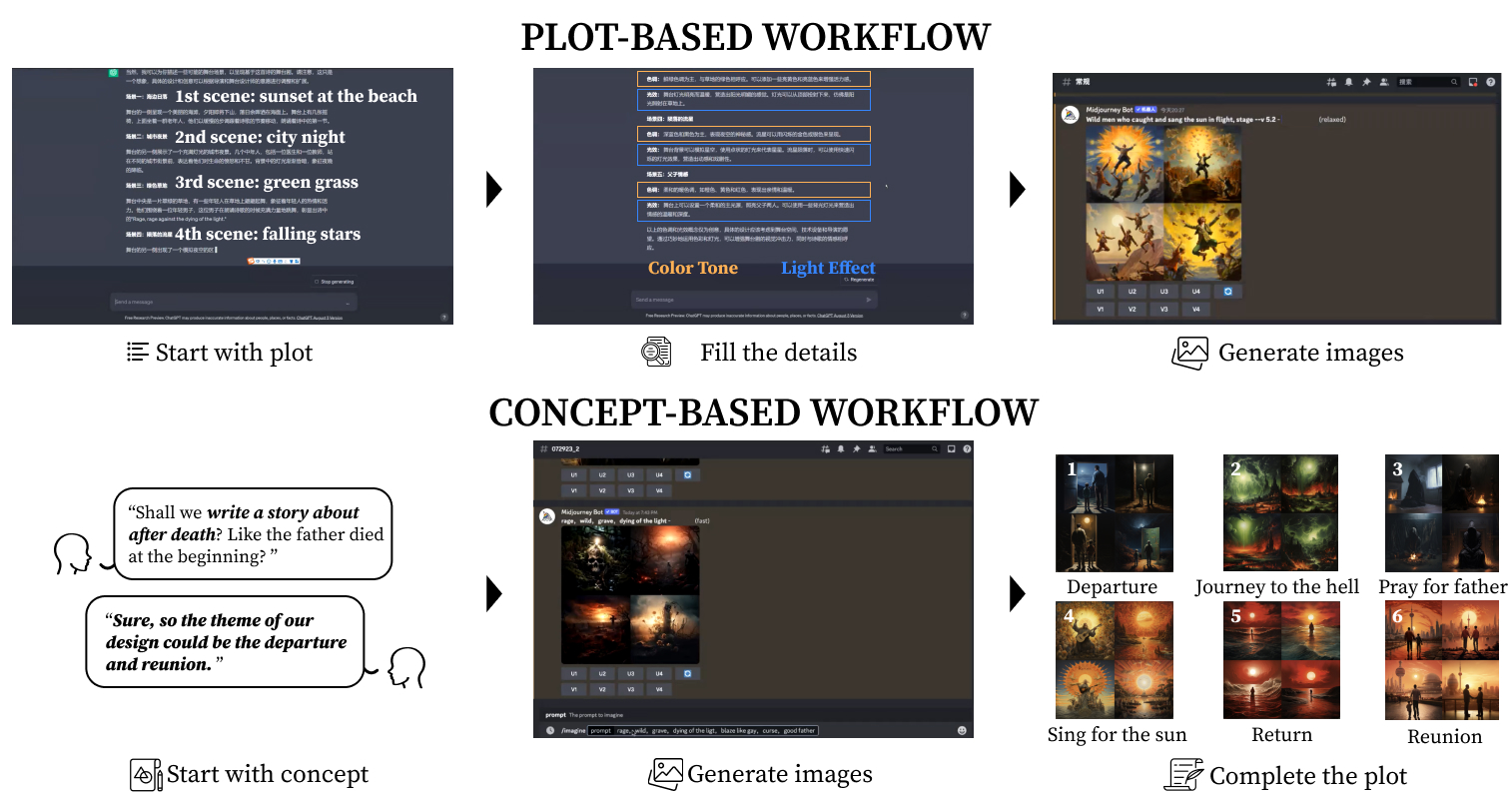}
    \centering
    \caption{Two overall strategies observed during the workshops: (1) Plot-based workflow: the groups started from imagining the plot for stage, and then filled up the details and generated images in the following process (this workflow is employed by group 2, group 3, group 8 and group 9). (2) Concept-based workflow: the groups started from building up the concept for stage, and then generated images and completed the plot in the following process (this workflow is employed by group 1, group 4, group 5, group 6 and group 7).}
    \label{fig:Overall co-prompting strategies}
\end{figure*}
\captionsetup{font=footnotesize, labelfont=bf, textfont=normalfont}

\subsection{Data Analysis}
\subsubsection{Semi-structured interview}
The qualitative data obtained from semi-structured interviews were subjected to a systematic coding process using an inductive approach \cite{braun2006using, saldana2015coding}. Initially, two researchers independently familiarized themselves with the data by re-reading the interview transcripts \cite{jones2020exploring}. This process allowed them to immerse themselves in the participants' responses and gain an understanding of the content. Following data familiarization, the researchers employed open coding to identify and label meaningful units of information within the transcripts \cite{braun2006using}. Two authors who are native Chinese speakers independently conducted coding of the transcripts. They individually analyzed the data and identified codes based on the content of the transcripts. Subsequently, they met to discuss any disagreements and worked towards reaching a consensus on the codes. To ensure consistency and enhance the reliability of the coding process, regular meetings were held between the researchers to discuss and compare their coding decisions \cite{miles2014qualitative}. Codes were assigned to phrases, sentences, or paragraphs that captured key ideas, experiences, or concepts related to the research questions. The coding process was iterative, with the researchers continually refining and revising codes as new insights emerged \cite{saldana2015coding}. Any ambiguities or uncertainties were further clarified by referring back to the original data.

After the initial coding phase, the researchers engaged in a collaborative process to identify potential themes from the generated codes. They reviewed the codes and searched for patterns, similarities, and differences amongst them \cite{guest2012applied}. Codes that shared commonalities were grouped together to form preliminary themes. This thematic grouping was guided by the content of the data and the research objectives. The identified potential themes were then critically reviewed and refined. The researchers examined the relationships between the codes within each theme, ensuring that they were coherent and representative of the data. Themes were revised or combined when necessary to accurately capture the underlying patterns and concepts present in the qualitative data. Throughout the coding process, researchers maintained detailed documentation of their coding decisions, including memos and reflective notes \cite{elo2008qualitative}. This documentation facilitated the traceability of the analysis, to maintain the rigor of the qualitative analysis.

\subsubsection{Co-prompting process}
A user journey map was employed as a data analysis method to gain insights into the experiences and interactions of a group of individuals while engaging with ChatGPT and MidJourney. One researcher reviewed all the recordings to collect all the prompting details during the experiment. Prompting processes were presented visually. The user journey analysis involved systematically examining the different touchpoints and stages that users went through during their interaction with GenAI systems. By mapping out the user journey, researchers were able to identify key moments of interaction, challenges faced, and valuable insights gained during the user's journey from text-based prompts to generated images \cite{stickdorn2011service, snyder2003paper}.

\section{Results}\label{sec:Results}
\begin{figure*}[ht]
    \centering
    \includegraphics[width=0.7\linewidth]{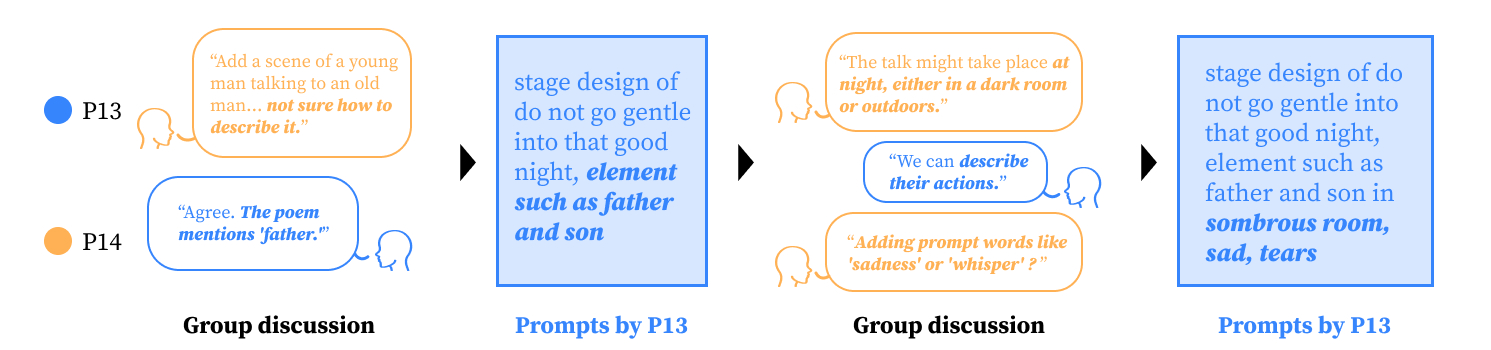}
    \caption{This figure illustrates the discussion between participants P13 and P14 during the conceptualization phase of the design. They did not have a clear and detailed design concept, but they could use GenAI to refine the details based on fragmented ideas to move forward.}
    \Description{This figure illustrates the discussion between participants P13 and P14 during the conceptualization phase of the design. They did not have a clear and detailed design concept, but they could use Midjourney to refine the details based on fragmented ideas to move forward.}
    \label{fig:Conceptual stage}
\end{figure*}

\subsection{Overall Strategies}
We thematically categorized the overall strategies adopted by participants into two distinct groups based on their workflow (Figure \ref{fig:Overall co-prompting strategies}). The first one, consisting of Group 2, Group 3, Group 8, and Group 9, adhered to a plot-based workflow. This approach involved first creation of a comprehensive narrative, with participants collaborating closely to construct a detailed plot-line for the performance. The subsequent stage involved elaborating on the specifics of the stage scenes based on the established narrative. Conversely, the remaining groups - Group 1, Group 4, Group 5, Group 6, and Group 7 - employed a concept-based workflow. Here, participants initially concentrated on defining the thematic elements, style, or central imagery for the stage, utilizing these foundational components as a springboard to craft the performance story-line and delineate the final scenes.

These two workflows seemed to be related to the professional backgrounds of the participants, and ChatGPT seemed to be used more  in Plot-Based groups during the collaboration process compared with Concept-Based groups. As demonstrated in the following sections, we observed an interplay between co-prompting and collaboration throughout the design process.

\begin{figure*}[ht]
    \centering
    \includegraphics[width=0.7\linewidth]{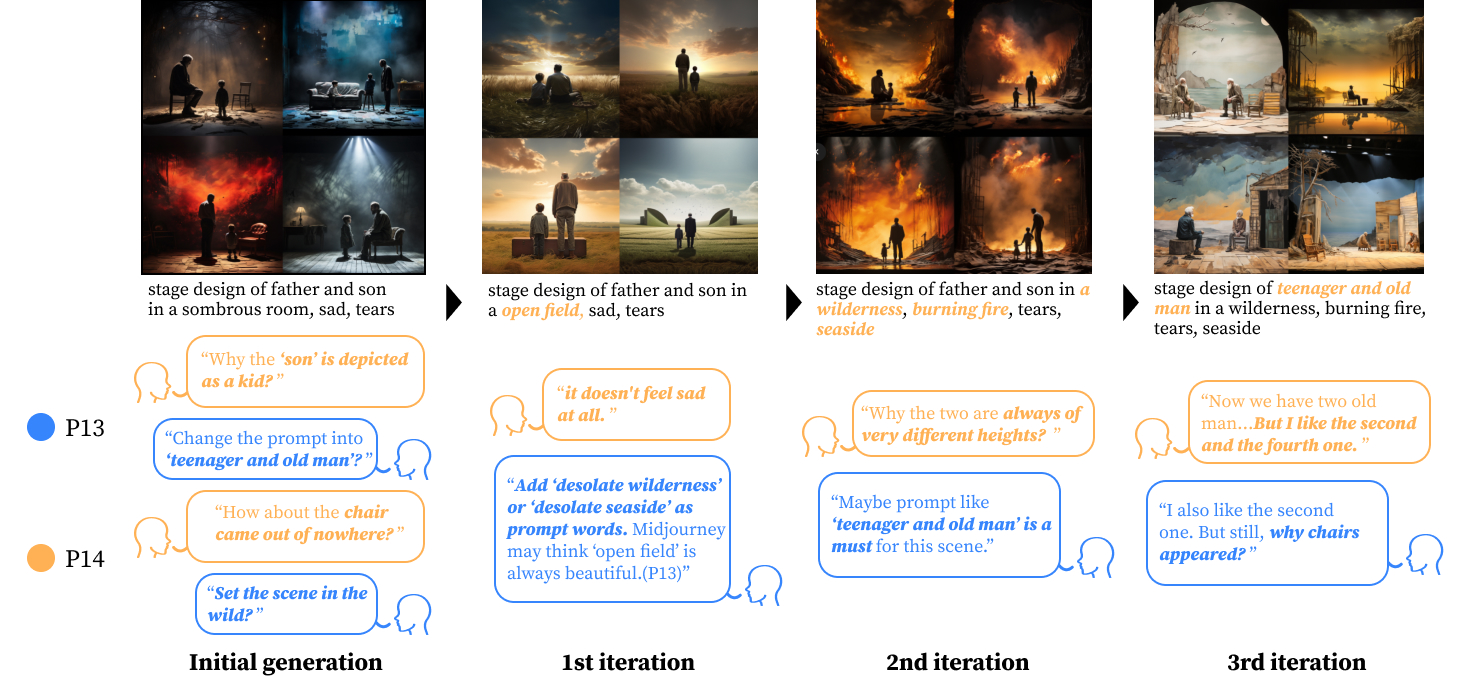}
    \caption{P13 and P14 modified the prompts together through discussions and iterations. The two participants wanted a scene with an elderly father and his son, a young man, standing together with a sad atmosphere, but the GenAI-generated results always depicted the son as a child rather than a young man, and the atmosphere lacked the feeling of sadness.}
    \Description{This figure shows the process of revising the prompt together for two participants, P13 and P14. The two participants wanted a scene with an elderly father and his son, a young man, standing together with a sad atmosphere, but the AI-generated results always depicted the son as a child rather than a young man, and the atmosphere creation sometimes lacked the feeling of sadness.}
    \label{fig: modify prompt P13P14}
\end{figure*}

\subsection{Collaboration Influences Co-prompting}

When collaborating together, participants felt that GenAI could swiftly produce content representing their ideas, thus quickly creating visuals of different concepts. Participants sought their collaborators' opinions in co-prompting, indicating that collaboration may affect the way participants decide on prompt words. This opinion exchange included discussions about prompt words and the sharing of ideas about generated images.

\subsubsection{Co-prompting to develop vague ideas into prompt words}
The co-prompting activities appeared to provide space for participants to collaboratively construct prompts, especially for some ideas that are still in the conceptual stage and challenging to articulate clearly. For example, we observed participants collaboratively fleshing out their thoughts through discussions regarding ideas from the poem that at first appeared to be simply inspiration ("sombrous", "sad") and developing them into concrete prompts ("dark room," "whisper") (Figure \ref{fig:Conceptual stage}).

\subsubsection{Co-prompting to overcome difficulties in adjusting the prompt}

It is known that participants can anticipate that GenAI will be able to comprehend their intentions in a manner similar to human understanding \cite{userprompting}. In our study, participants highlighted the challenges associated with making GenAI fully grasp their ideas, noting that it generally interprets only the literal meaning of the prompts, often failing to discern the underlying intentions or emotional nuances. P6 underscored this difficulty, stating, \textit{"To utilize the AI properly, one needs to understand the correct methodology for instructing it to produce the desired results."} Participants found GenAI's limited capacity for association to be a hindrance, necessitating numerous iterations of clarifying and modifying their prompts to convey their ideas accurately.

To overcome the difficulties in articulating ideas to generative AI, participants collaboratively modified the prompt words, urging GenAI to deliver satisfactory output. They also speculated on how Midjourney interprets their prompts. When P13 and P14 prompted Midjourney to generate an image depicting a father and son in a sad atmosphere, they were dissatisfied with the generated results, so they tried refining their prompt words together (Figure \ref{fig: modify prompt P13P14}).


\subsubsection{Communication with AI can still be difficult even with the help of others}
\begin{figure*}[ht]
    \centering
    \includegraphics[width=0.7\linewidth]{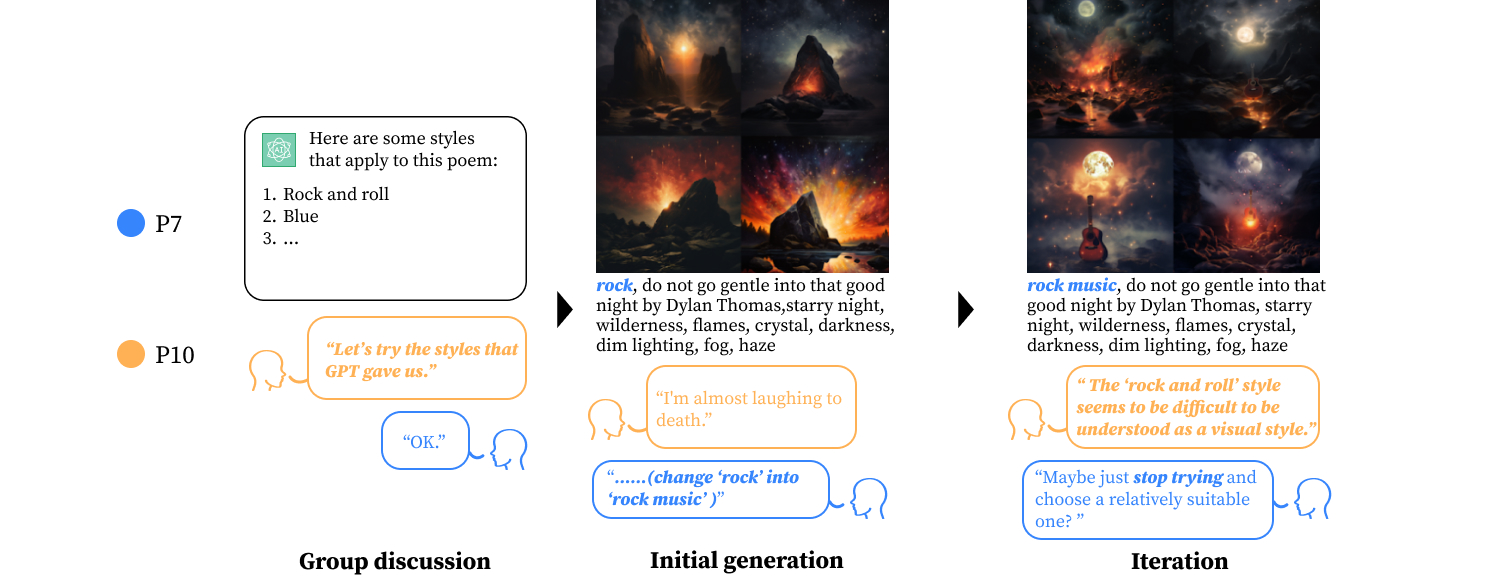}
    \caption{P7 and P10 tried to give the image “rock and roll” style by prompt 'Rock', but the Midjourney interpreted 'Rock' as a stone in the generated image. After trying to change it to 'Rock music', guitars appeared in the picture generated by GenAI. At this point the participants gave up trying because the GenAI could not understand their idea.}
    \Description{This figure shows the Midjourney prompting process for P7 and P10. Two participants tried to give the image with “rock and roll” style by prompt 'Rock', but the Midjourney interpreted 'Rock' as a stone in the generated image. After trying to change it to 'Rock music', guitars appeared in the picture generated by AI. At this point the participant gave up trying because the AI could not understand their idea.}
    \label{fig: P7P10}
\end{figure*}

Participants reported difficulty in communicating in a way that both human collaborators and GenAI could understand one another, highlighting the challenges posed by iterating prompts precisely within a group. Prompting the AI proved to be a particularly burdensome task that often discouraged further attempts at iteration. P7 reflected, \textit{"The experience was discouraging, and I found myself tempted to abandon the task. The GenAI's lack of associative abilities made communication both draining and time-consuming, necessitating repeated clarification of our initial inputs."} (Figure \ref{fig: P7P10}). We also observed that participants would abandon a generation attempt after a few unsuccessful trials if they still could not achieve a satisfactory result. Moreover, participants noted a heightened sense of caution when prompting in a collaborative context to prevent errors that could potentially decrease efficiency and let the group down. P8 emphasized, \textit{"In a co-prompting setting with others, I exercised greater caution to avoid mistakes and maintain efficiency."}


\begin{figure*}[ht]
    \centering
    \includegraphics[width=0.7\linewidth]{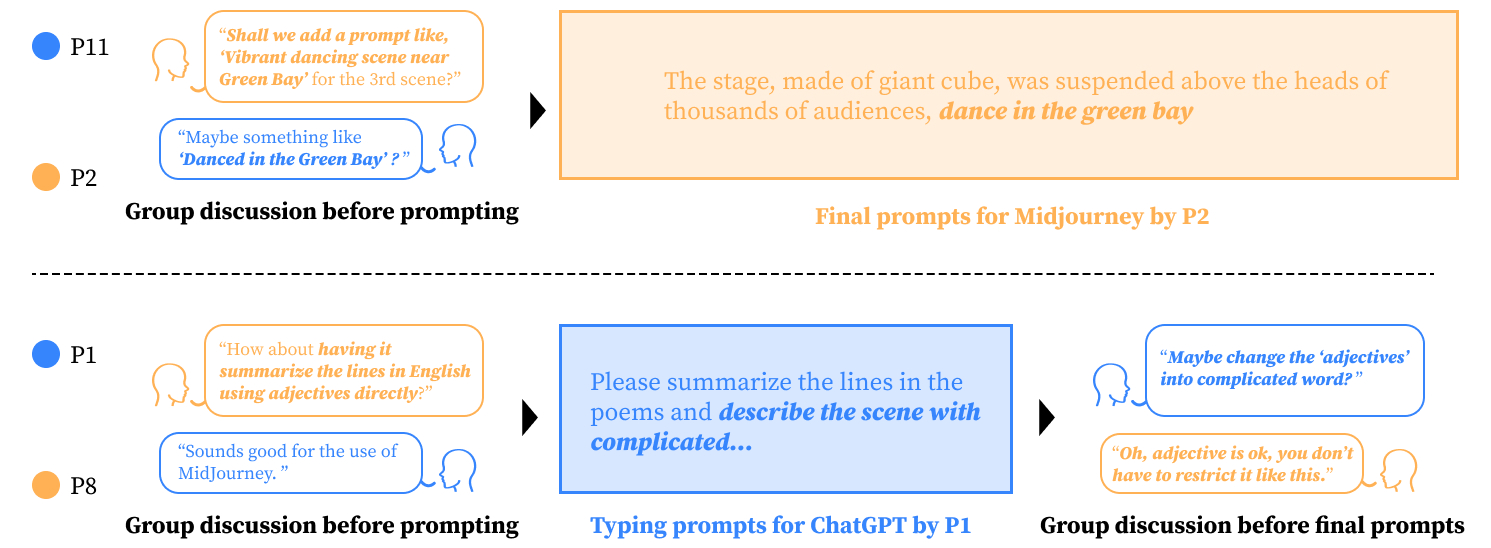}
    \caption{Discussion before prompting. (Above) P2 asked P11 for specific advice on prompt words before he actually typing them into Midjourney. (Below) P1 asked P8 for further advice before changing the prompts based on his own ideas. Both P2 and P1 asked their teammates for their thoughts and opinions before finalizing the prompt.}
    \Description{This figure shows the Midjourney prompting process for P2 and P11, and the ChatGPT prompting process for P1 and P8. Both participants, P2 and P1, asked their teammates for their thoughts and opinions before finalizing the prompt.}
    \label{fig:ask teammate 1}
\end{figure*}

\subsection{Co-prompting Influences Collaboration}

\subsubsection{Co-prompting inspires discussion and leads to testing of ideas}
Participants reported using GenAI to quickly generate content replete with detailed representations, enabling participants to test their ideas in rapid succession and at low labor costs compared with manual production. This efficiency appears to encourage participants to explore and iterate their ideas with diminished psychological burden. P3 noted, \textit{"The manual drawing process is slow and laborious, particularly when revisions are necessary. GenAI facilitates easy adjustments, thereby fostering a richer ideation process."} This sentiment was echoed by P6, who emphasized the speed of Midjourney in image generation, allowing for \textit{"low-cost trials and errors, and facilitating uninhibited idea validation."} Furthermore, P7 acknowledged the neutral position of GenAI in the collaborative setting, remarking, \textit{"GenAI serves as a neutral third party on our team... we don't need to worry about its feelings."} Thus, the incorporation of GenAI into the team appears to encourage participants to experiment and iterate their ideas, aiding individuals in materializing and refining their concepts to achieve favorable results.


\subsubsection{Participants wanted to hear more from their teammates in co-prompting}
\begin{figure*}[ht]
    \centering
    \includegraphics[width=0.7\linewidth]{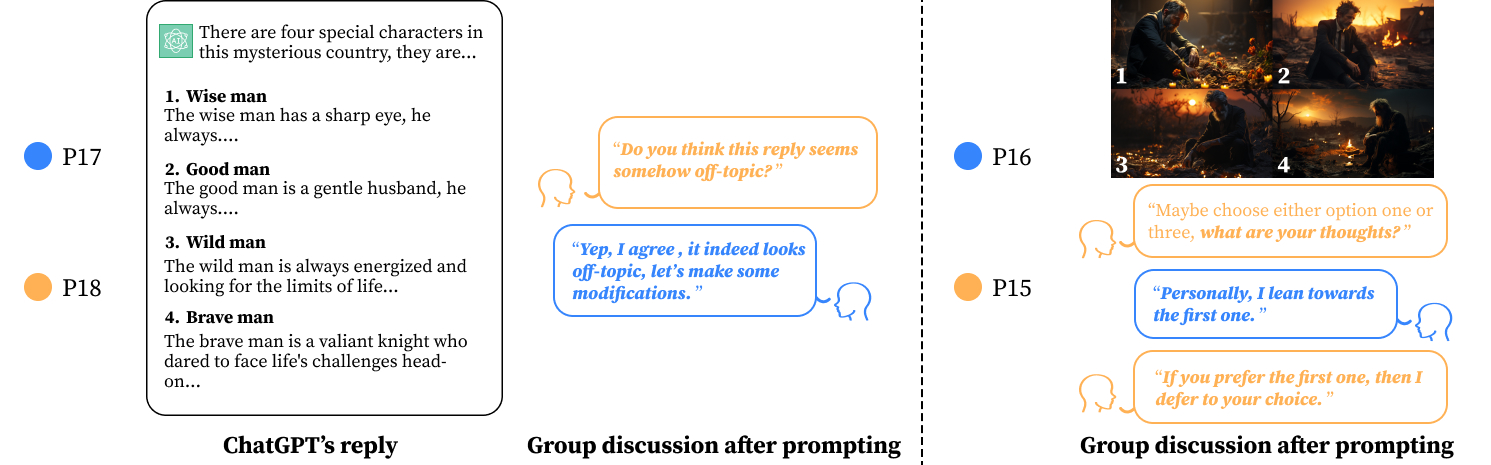}
    \caption{{Discussion after seeing results. (Left) P18 asked P17 about his ideas concerning ChatGPT's reply since P18 seemed not satisfied with it. (Right) P15 asked P16 about P16's preference when selecting the images, and they eventually chose the one that P16 felt good about. Both P18 and P15 took the initiative to ask their teammates for their thoughts after the results were seen.}}
    \Description{This figure illustrates the ChatGPT prompting process for P17 and P18 as well as the Midjourney prompting process for P15 and P16. Both participants, P18 and P15, took the initiative to ask their teammates for their thoughts and opinions after the results were generated.}
    \label{fig:ask teammate 2}
\end{figure*}

\begin{figure*}[ht]
    \centering
    \includegraphics[width=0.7\linewidth]{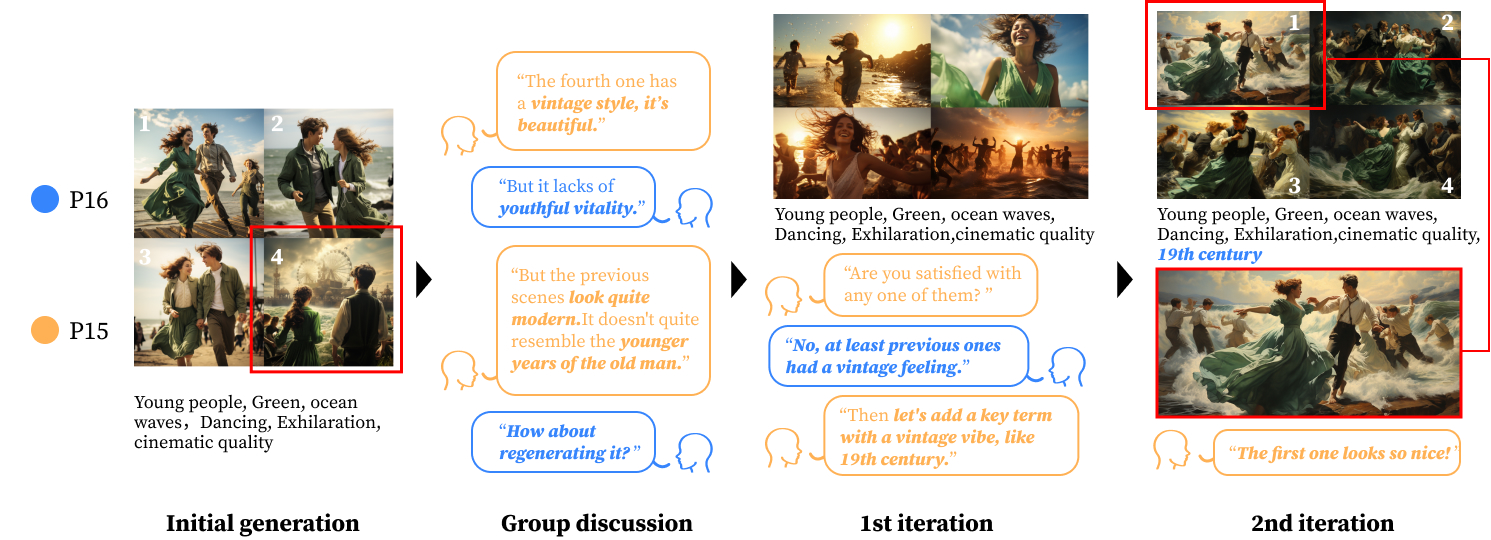}
    \caption{Iteration until mutual satifaction. The prompting process for P15 and P16 showed them dissatisfied with the generated content. Through continuous discussion, they determined how to modify the prompt iteratively until both of them were satisfied with the results.}
    \Description{This figure illustrates the Midjourney prompting process for P15 and P16. Both participants were dissatisfied with the generated content, and through continuous discussion, they determined how to modify the prompt and finalize the desired image.}
    \label{fig:Modify prompt together}
\end{figure*}

During the prompting process, participants consistently sought to involve their collaborators, soliciting suggestions and encouraging them to actively engage in evaluating and modifying prompts collaboratively. This was observed at various stages of the process, both in the initial stages of crafting the prompt (Figure \ref{fig:ask teammate 1}) and following the generated output phase (Figure \ref{fig:ask teammate 2}). Throughout the workshop, participants undertook to further refine and optimize prompts based on the feedback received from their peers. This iterative collaborative process could cultivate outcomes that were met with satisfaction by all team members Figure \ref{fig:Modify prompt together}).

\begin{figure*}[ht]
    \centering
    \includegraphics[width=0.7\linewidth]{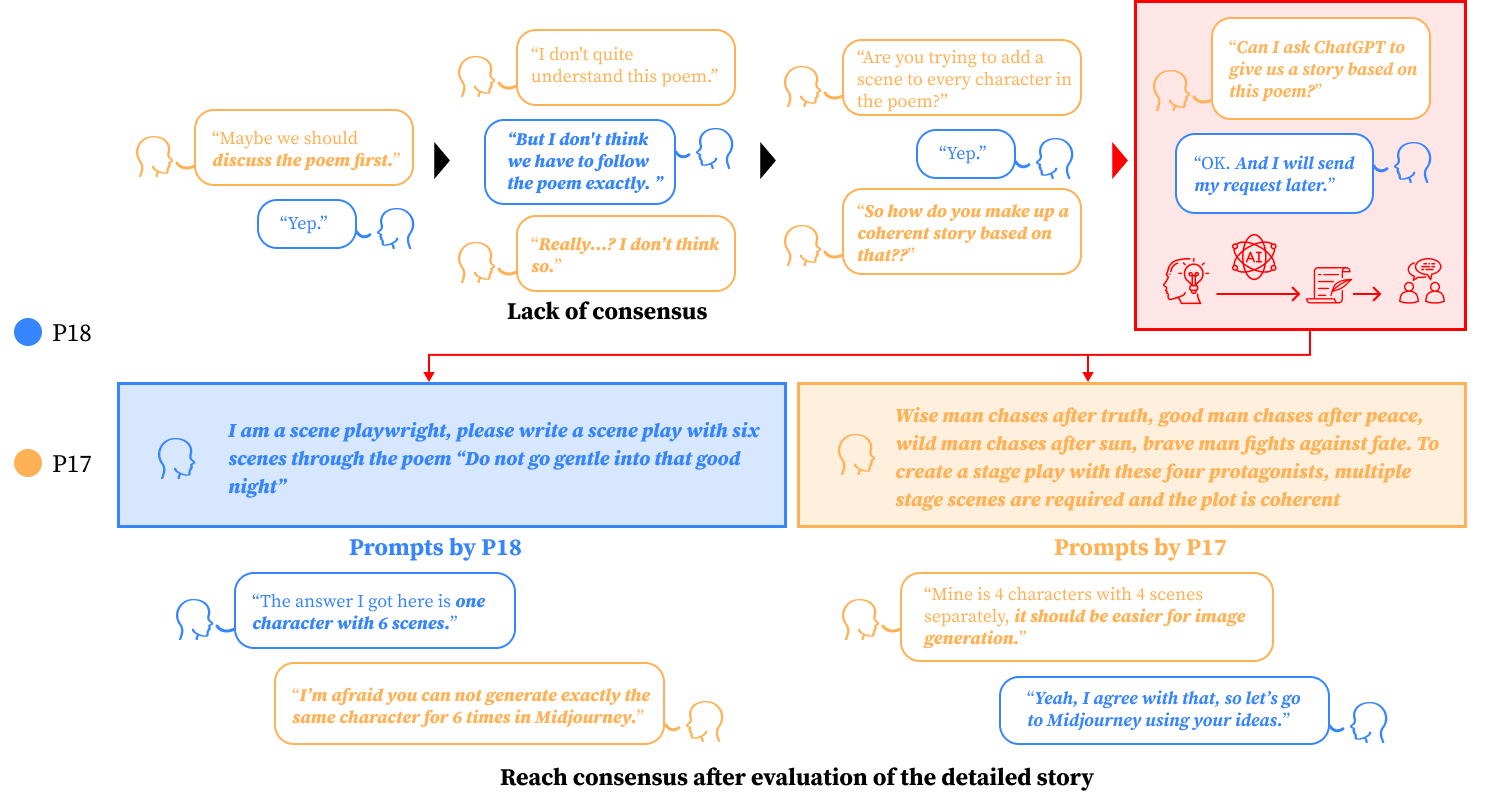}
    \caption{The prompting process for P17 and P18 showed conflicting ideas in design ideas in the early stage of cooperation and had difficulty convincing each other through communication (upper). By using ChatGPT to convert their ideas into concrete text descriptions and present them to each other, P17 and P18 finally reaching a consensus through discussion on the generated outcomes.}
    \Description{This figure illustrates the ChatGPT prompting process for P17 and P18. The upper dashed box shows the process of cooperation between two participants who had plenty of conflicts in design ideas in the early stage of cooperation and had difficulty convincing each other through communication. The lower box shows the process of the two participants using ChatGPT to convert their ideas into concrete text descriptions and present them to each other, and finally reaching a consensus through the following discussion on the generated outcomes.}
    \label{fig:GPT validate}
\end{figure*}
\subsubsection{Co-prompting can enhance mutual understanding in teams}
We found that prompting GenAI collaboratively appears to enhances the way participants convey specific information, facilitating the communication of concepts that might be challenging to express verbally. The end result is the materialization of abstract thoughts, contributing to understanding amongst human collaborators. Although GenAI sometimes gave rise to discrepancies in ideas from the original concept, participants reported that GenAI's contribution was beneficial in effectively conveying their core ideas. As P8 stated, \textit{"While the information provided by AI wasn't fully aligned with my idea, it was sufficient to convey my thoughts."} In this context, GenAI appears to serve as a bridge, enabling a more tangible representation of participants' ideas and fostering mutual understanding in teams.

Participants found that co-prompting allows for discussions where ideas can be shared and assessed based on GenAI-created representations. P13 recalled an instance during her collaboration when her collaborator's interpretation of the GenAI-generated output changed her view: \textit{"When we used Midjourney to generate this set of images, I initially thought the atmosphere did not align with the tone I interpreted from the poem. However, [P14]'s insights into the first image made me reconsider. Had I been working alone, I would probably have dismissed the set of images."} Moreover, in situations of divergent views, the GenAI-created outputs seemed to become a basis for negotiation, offering more concrete understanding of each participant's perspective, and subsequently easing the path to reaching a consensus. P17 and P18, for instance, utilized ChatGPT to transform their individual ideas into stories. They then used these narratives as a means to deliberate on the feasibility of their ideas, a dynamic that increased mutual understanding and played a role in achieving a harmonious solution. (Figure \ref{fig:GPT validate}).

\begin{figure*}[htbp]
    \centering
    \includegraphics[width=0.7\linewidth]{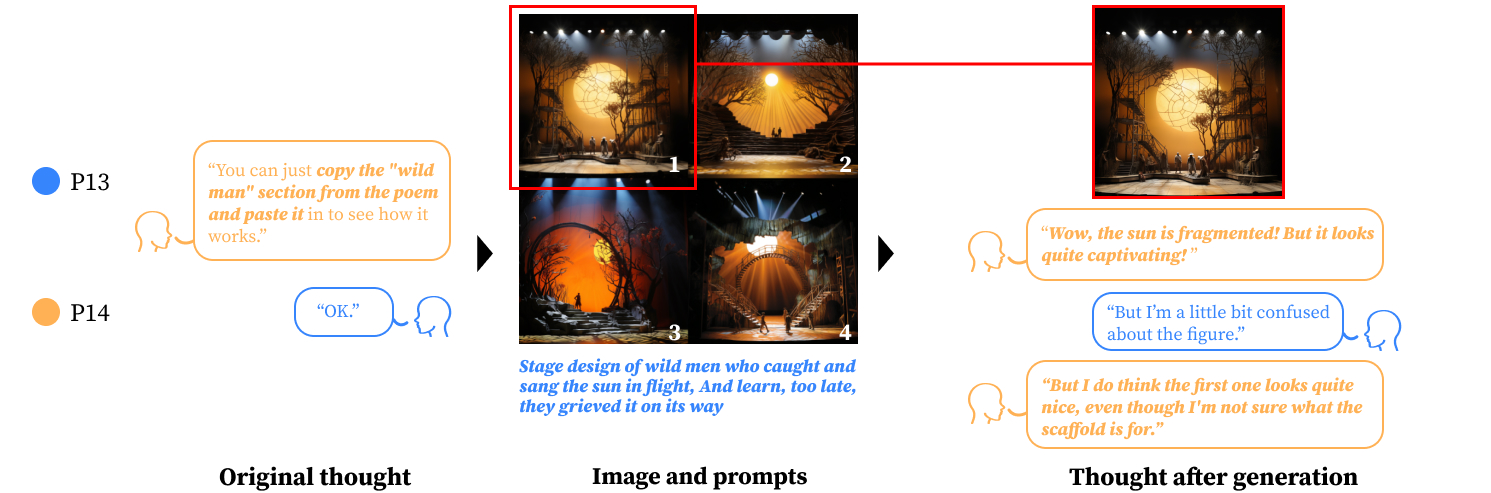}
    \caption{P13 and P14 were surprised by the 'broken moon' generated by GenAI since it was not described in the prompts, but they seemed very satisfied with this unexpected result, and chose this picture in the final submissions.}
    \Description{This figure illustrates the Midjourney prompting process for P13 and P14. Two participants got an unexpected picture after entering the prompt, but they turned out quite like the unexpected image and finally chose this picture in the final submissions.}
    \label{fig: broken moon}
\end{figure*}
\begin{figure*}[htbp]
    \centering
    \includegraphics[width=0.7\linewidth]{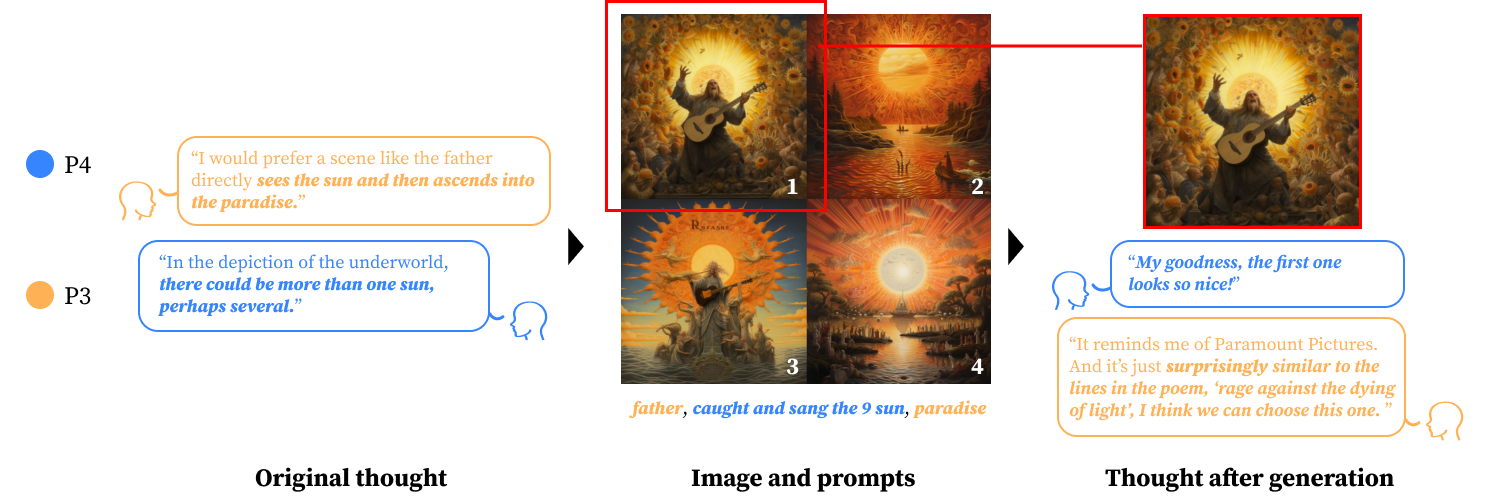}
    \caption{P3 and P4 were surprised by the generated images that Midjourney gave them. At first, they did not understand why there was a deviation of the generated image from the initial idea, but after discussion, they realized that the image could also form a good connection with the design theme, because it matched with the lines in the poem. Finally, they chose this image as one of the final works.}
    \Description{This figure shows the Midjourney prompting process for P3 and P4. The two participants got an unexpected picture after entering the prompt with an already-formed idea. At first, they did not understand why there was a deviation of the generated image from the initial idea, but after discussion, they realized that the image could also form a good connection with the design theme, and finally chose this image as one of the final works.}
    \label{fig: unexpected P3P4}
\end{figure*}

\subsection{Participants’ Perception Towards GenAI and Humans in Collaboration}
During the interviews, participants were asked to describe their perception of GenAI in the process. Participants tended to view GenAI as helpful agents assisting them with tedious work. Participants selected GenAI-created outputs that aligned with their ideas while also expressing a desire for GenAI to provide them with unexpected inspiration. Although participants prompted GenAI to do creative work, they voiced their concern that human participants' creativity might be distrupted by GenAI workflows.

\subsubsection{Human is responsible for guiding the GenAI to fulfill the laborious task}
We observed that several groups chose to assign creative responsibilities to GenAI, taking on roles similar to curators who oversee, critique, and choose the GenAI-created outputs. The majority perceived GenAI as a tool rather than a collaborative partner, attributing this to the GenAI's incapability to actively engage in human discussions, comprehend emotional nuances, and partake in decision-making processes within the team. In contrast, P9 personified GenAI as a vital team member, noting its capacity to enhance productivity and foster creativity throughout the workshop. This sentiment was echoed by P17, who likened the GenAI to a consultant colleague offering valuable insights when humans got stuck coming up with ideas.

During interviews, participants described themselves in various roles, such as translators (P8), project managers (P8), film producers (P7), commissioners (P7), and design instructors (P7). They perceived GenAI through different lenses as well, categorizing them as designers (P8), listeners (P13), secretaries (P2), teammates (P9), service providers (P7), and design students (P7). Indeed, irrespective of the categorization, a consistent theme emerged: participants viewed GenAI as responsible for content generation and laborious tasks, while they steered the direction and provided guidance.

In particular, participants tended to view GenAI as a helpful agent that assists them, believing that it increases efficiency and saves labor in their collaborative work. They delegated tasks to GenAI over time, showing a shift in the distribution of labor. For example, after discussing their interpretation of the poem, P7 and P10 decided to use ChatGPT to extract pivotal words for further prompting. P7 explained, \textit{"We used ChatGPT to save time and avoid having to read and interpret the poem over and over again. For me, GenAI is a time-saving and labor-saving tool."} Further instances of ChatGPT utilization encompass translation endeavors, materials organization, and the extraction of design recommendations from pre-existing ideas or content.

\subsubsection{Participants reported a preference for GenAI's output to match with their own, but also welcomed unexpected inspirations}
After co-prompting GenAI, participants evaluated the outcome based on their preferences. They recognized the aesthetic quality of the GenAI output, interpreted the output, and selected the best matches to their expectations, iteratively revising the prompts. \textit{"We need to steer the GenAI in generating images and emphasize certain keywords.”} (P7).

However, participants also saw the GenAI output could provide them with unforeseen inspiration. "\textit{You develop ideas based on your own experiences, which can lead you in a certain direction; AI could introduce you to a new way of thinking.}” (P8). \textit{"I think the GenAI has its own preference due to the data they are trained, so the output generated by AI could offer us new inspirations."} (P17). Participants were delighted to accept ideas offered by GenAI. For instance, when Midjourney produced an image featuring a 'broken moon' (Figure \ref{fig: broken moon}), P14 expressed her surprise: "\textit{When we think of the moon, we usually imagine a round, glowing orb. GenAI's depiction of a broken moon was unexpected and intriguing, it changed the atmosphere in a positive way. I appreciate these subtle touches}." Participants were willing to embrace unexpected content from GenAI, even when it did not exactly represent their own ideas.

However, when GenAI failed to deliver the unexpected insights participants hoped for, disappointment can set in. \textit{"I wish GenAI could provide me with more 'aha moments'—something beyond what my prompts specify."}(P3) (Figure \ref{fig: unexpected P3P4}). He added that if GenAI could not provide additional inspiration, he would opt not to use it.

\subsubsection{Participants perceived the original ideas of humans to be at a higher category than those of GenAI}

Numerous participants with backgrounds in design underscored concerns about the potential diminishing of originality and creativity when incorporating GenAI into collaborative creative tasks. They at times expressed a preference for maintaining independence in their thinking devoid of GenAI. Echoing this sentiment, P1 expressed the desire to express individual emotion in creative endeavors, emphasizing, \textit{"If I participate in the workshop again, I would want a human to interpret the poem, come up with subjective feelings. ChatGPT is powerful, but what it generated cannot represent my feelings."}

Furthermore, those well-versed in stage design voiced skepticism regarding GenAIs’ depth of understanding in the nuanced field of stage design. P1 found ChatGPT’s insights into stage design somewhat rudimentary, while P14 criticized Midjourney’s rigid approach to the discipline. Moreover, participants showed a tendency to value ideas from their human counterparts over suggestions given by the GenAI. Participants would invite their teammates to comment on and refine the generative outcomes instead of asking GenAI. This collaborative approach often led to more diverse results. P15 observed, "Numerous suggestions from my teammates translated into outputs that were surprising and innovative."

Participants tended to trust their human collaborator more than the generative AI. They invited their collaborator to determine the validity of GenAI’s output and trusted GenAI’s output only if it aligned with human ideas, even if they were surprising ones. P17 mentioned, \textit{"If I could find someone to check the output from GenAI with me, I would think the results given by GenAI are more reliable."} Some participants expressed appreciation for in-depth discussions on creative content, a facet that was somewhat overshadowed by the task emphasizing co-prompting strategies. Echoing this, P1 lamented, \textit{"In the workshop, we paid excessive attention to how to use the tool and had less conversation on the creative content."}

\section{Discussion}\label{sec:Discussion}
\subsection{RQ1: Challenges for Teams in Co-Prompting GenAI for Creative Design and Strategies for Overcoming Them}

Co-prompting appears to be a double-edged sword for creative collaboration. Co-prompting enhances mutual understanding between human collaborators and enables individuals to get help from their human counterparts. However, prompting generative AI remains challenging for novice users. Moreover, introducing GenAI in collaboration may induce participants to pay more attention to GenAI rather than to the content being created.

\subsubsection{Prompting remains challenging in a collaborative setting}
During the intervention, participants encountered difficulties in verbalizing their thoughts, particularly in ensuring that both GenAI and their human collaborators understood their intentions. The incorporation of AI in the collaboration added complexity to the communication. Furthermore, participants struggled to comprehend how GenAI would interpret prompts, particularly in text-to-image generation. At times, participants felt that GenAI grasped only the literal meaning of prompts, failing to capture the nuanced emotions and context they intended. This led participants to adjust their prompts to enable GenAI to understand their ideas.

Our findings indicate that participants are still applying a mental model of human-human interaction when interacting with AI, particularly with text-to-image AI. This observation aligns with previous findings on why non-AI expert users face challenges when prompting Large Language Models \cite{userprompting}. Participants expected the AI to understand the context of their prompts, such as interpreting "rock" as "rock and roll" rather than an actual rock. However, the text-to-image language model may interpret users' prompts more straightforwardly, without inferring contextual or metaphorical semantics that humans might naturally apply. This misperception towards the text-to-image model may lead participants to rephrase their prompts several times to achieve a preferable generation result, resulting in frustration after receiving undesirable outcomes (4.2.3). While iteration is considered a common strategy in prompt engineering \cite{Designguidance}, the repeated prompting by non-expert users in a collaboration context can be frustrating due to the lack of transparency in GenAI's mechanism.

Additionally, participants appeared to struggle with adjusting the prompt words, which may explain why some participants felt they were too focused on figuring out how to better use the GenAI rather than exchanging their creative ideas and having more in-depth discussions on them (4.4.3). In previous research, when the music composing process was intervened by GenAI, the participants also felt the GenAI might hinder the depth of their collaboration. They switched their focus to how to improve the content made by AI instead of having more creative engagement \cite{suh_ai_2021}. Moreover, research suggests that a human-AI hybrid team has worse performance than a human-only team because the contribution of AI lowers participants' mental demand and gives them an illusion of success, making the participants put less effort into the task \cite{zhang_cautionary_2021}. The participation of AI in the co-creation process could distract participants from their creative interactions with team members, shifting their focus to how to prompt properly (4.4.3). Additionally, the output given by the GenAI may satisfy participants in terms of creating outcomes, making them less diligent in ideating for their own creations in the design process.

In summary, working with co-prompting in teams appears to be a double-edged sword, facilitating creative ideation on the one hand to support team collaboration, but adding mental demands of strategic prompting which could reduce team performance.


\subsubsection{Strategies in collaboration could enhance prompting}
Throughout the intervention, participants employed various strategies for prompting GenAI, including collaborative structuring of prompts and actively offering suggestions. 
Participants also strategically assigned laborious tasks to GenAI. This approach allowed them to quickly test out their ideas and use the generated outputs to materialize their abstract concepts, aiding in mutual understanding amongst human collaborators.

The collaboration enabled the participants to seek help in prompting from their human collaborators (4.2.1 and 4.2.2). Previous research has found that when non-expert users prompt GenAI tools individually, they seek help from online resources \cite{userprompting}. Creative also actively look for unique prompt words from various resources, suggesting that constructing prompt words may require external assistance. The human participants in the workshop could serve as an "outer party" for their collaborators to request help.

Co-prompting is beneficial not only for enhancing the prompting process but also for enriching the collaborative experience. As one study notes, individuals tend to employ AI to handle large volumes of work when they emphasize productivity, thus reducing their own workload \cite{biermann_tool_2022}. In the co-prompting setting, participants prompted GenAI to reduce their workload and save time in the team, assigning the more burdensome tasks to the GenAI. Furthermore, participants utilized the outputs from GenAI as a visual medium to communicate their abstract ideas, facilitating mutual understanding between collaborators (4.3.3). Previous research has found that visual representation can reduce misunderstanding and conflict in collaboration \cite{NGUYEN_systematic_2022}. In co-prompting, participants applied strategies that utilize GenAI to not only lighten their task load but also enhance their efficiency in communicating ideas.

\subsection{RQ2: Perception towards the roles of GenAI and human collaborators in co-prompting}

Our findings suggest that participants expect GenAI's output to fully represent their ideas, while they also welcomed the unexpected inspiration delivered. This may be due to participants' varying expectations of the roles of GenAI. Previous work has found that creative writers preferred to retain control over their writing strategies while co-creating with GenAI \cite{biermann_tool_2022}.We might expect designers to have similar expectations of GenAI when co-prompting with other designers, with GenAI respecting and adhering to the team's strategies in the design workflow. Indeed, our findings show that participants assigned GenAI to a subservient role in the design process (4.4.1) and categorized their strategies as more valuable than GenAI's (4.4.3), consistent with the findings on creative writers. In contrast, a study on designers suggested that they are more tolerant of GenAI's output accuracy when they view GenAI as a tool for inspiration \cite{KWON_understanding_2023}, indicating a trade-off between expecting GenAI's support in a subservient role and its inspirational role in ideation. It appears that designers evaluate the ability of GenAI depending on the purpose for using it. Our study suggests that the role of the other human in the system falls into a hierarchy above the GenAI (4.4.3) even in the inspiration-generation phase. 

Another concern we uncovered was participants' worry about the loss of creativity and lack of depth in thinking due to GenAI's intervention in teamwork, as expressed during the interviews. Previous research suggested that GenAI can lower the mental demand of a high-performing human design team and make them explore less in a design task \cite{zhang_cautionary_2021}. In our study, participants tended to follow the suggestions delivered by GenAI and missed chances to explore their design task further, which led participants to reflect during the post-workshop interviews that they should not rely on the content provided by GenAI. Also, previous work showed that writers are more willing to collaborate with GenAI when they lack confidence \cite{biermann_tool_2022}. Most of the participants did not have extensive experience in working in the discipline of stage design, which may have caused them to lack confidence when working on design tasks, and hence more receptive to suggestions from GenAI. 

We observed participants utilizing plot-based and concept-based workflows depending on their background. We noted that participants with design backgrounds are more likely to apply the concept-based workflow (4.1). However, the plot-based workflow we observed is closer to the actual workflow of stage designers: they should read the script first and then identify all elements needed for the stage \cite{stagedeisgn2}. Previous work has shown that undergraduates in architecture and industrial design are more likely than design Ph.D. candidates to not plan the design process in advance \cite{goldschmidt_design_2013}. Thus we speculate that participants' actions in the workshop could be influenced by their educational background.

One phenomenon we observed was participants' anthropomorphization of the GenAI (4.4.1). Some participants viewed GenAI in a more passive position (as a service provider), while others saw it in a more active role (as a teammate). Based on a previously published work, when individuals co-create with GenAI, they cycle between two states: highly engaging with GenAI and accepting its suggestions (Co-Creative Agentive Flow), or viewing GenAI as a tool to support their creation (Tool-Supported Creative Flow) \cite{WhenIsATool}. Most participants perceived GenAI as guided by humans to create content, indicating a predominant experience of Tool-Supported Creative Flow in the co-prompting process. Nonetheless, they also accepted unexpected inspiration from GenAI throughout the workshop, showing that Co-Creative Agentive Flow exists in the co-prompting process. We speculate that the workshop design led participants to rely more on Tool-Supported Creative Flow: GenAI only responded when participants collectively decided to prompt it, positioning GenAI passively from the start. This is also supported by a participant (P18) who considers GenAI a tool that cannot actively join the discussion and decision-making process in their team. Furthermore, our observations indicated that participants tended to trust their human collaborators more than GenAI (4.4.3) and sought to have deeper discussions on creative ideas with humans collaborators only (4.3.2), suggesting a preference for more creative engagement with human collaborators over GenAI.

\subsection{Design Implications}

To optimize GenAI's effectiveness in creative collaborations, future systems should facilitate more human-like interactions and ensure non-expert users have accurate perceptions of GenAI's capabilities. Thus, users can focus more on actively discussing their creative ideas with human collaborators instead of being sidetracked by the complexities of using GenAI. We also suggest adapting GenAI to different stages of the collaborative design process, such as brainstorming and testing. For instance, during brainstorming, GenAI could produce more varied outputs to spark inspiration, actively participating in discussions rather than merely responding when prompted. In the concept validation stage, GenAI should more precisely mirror users' intentions in the way they want to visualize their ideas, aiding them in effectively communicating to their human colleagues. A scenario-specific GenAI system would enhance team creativity and efficiency, guiding them through challenging moments while maintaining human ingenuity.

Furthermore, in designing future GenAI systems for collaborative creativity, we must consider the team members' performance to tailor interactions. Our work shows that less confident users are likely to embrace GenAI's suggestions, potentially leading to greater reliance on GenAI. However, users highly value their collaborators' ideas, so systems should offer users opportunities for reflection, aiming to decrease their dependency on GenAI.

\subsection{Limitations}
\subsubsection{Online collaboration setting}
One limitation is the exclusive use of an online collaboration setting, potentially influencing dynamics and outcomes compared to offline scenarios. When participants in our study worked with each other, usually one designer was the person entering the prompts while the other designer contributed. In real-world environments, this type of work is likely to have greater affordance for the other human when they are working offline together. This suggests that if the task is carried out offline, participants may value the human participants even further and limit the role of the GenAI to be a supporter. It is also possible that online and offline strategies are similar, given that one person would be in charge of the actual entering of prompt in both cases. However, we envision in offline cases that the prompt entering may be switched between the participants more frequently, giving the GenAI a greater data-entry and visualization role.

\subsubsection{Language and cultural nuances}
GenAI models we used for this study were trained on English language data, while participants were native Chinese speakers, even though they performed the prompting in English. This mismatch may introduce linguistic and cultural nuances impacting the interpretation and generation of GenAI content. Exploring AI models specifically trained on Chinese language data could address this limitation and answer questions about whether the primary language of humans and GenAI must match for effective engagement to occur. This also suggest that some of the differences in perception of the other humans and GenAI in our study could be due to subtler language issues.

\subsubsection{Lack of control study of interaction without GenAI}
Our task put humans in pairs with GenAI to co-prompt together. However, what would happen if no GenAI were present and the humans were simply designing as a team? Related works in collaborative design have shown that in such interdisciplinary contexts, trust and communication are key themes of varied styles of collaboration \cite{NGUYEN_systematic_2022}. However, probing what happens when humans collaborate with humans would somewhat reproduce existing literature, and does not tell us about the emerging case of AI-supported creativity. We do not claim that co-prompting is \textit{necessary} for any particular type of observed engagement, but rather that co-prompting can lead to certain types of engagement with, and perception of, GenAI. A study that does not use GenAI would not tell us how that engagement goes because it would not be studying a team's interaction with GenAI. Thus, we have focused here on patterns of interactions and perceptions that teamwork with GenAI can lead to.

\subsubsection{Homogeneity of participant characteristics}
Participants in the study shared similar demographic characteristics, such as age around 20 and being university students, limiting the generalizability of findings to a broader population. It is possible that those of older demographics without as much experience with technology would perceive GenAI agents differently, such as treating them as experts, or rather mistrusting them. Future research should aim to include participants with diverse backgrounds and expertise to obtain a more complete understanding across the population.

\subsubsection{Limited team size and generalizability} 
The small team size of two members may not fully capture the complexities of larger teams, limiting the generalizability of findings to larger teams or real-world settings with more individuals. We may find that when dealing with larger teams, the human collaboration element becomes much more intertwined and difficult to disentangle on its own, and that collaborative prompting may be relegated to one individual. Thus differences in the team size may either enhance or reduce the reliance on prompting of GenAI. A proper study of the subject would involve giving team tasks that rely on multiple parties to accomplish and allowing them to work with GenAI either separately or with a single system.

\subsubsection{Generalizability to other tasks} 
The use of a specific task (inspired by a poem to design for the stage) may not represent the range of tasks in human-AI teamwork. The findings and insights gained from our study should be interpreted within the context of a creative task that involves image generation based on textual inspiration. For example, a purely storytelling task may put the GenAI with a more equal footing with the human because they are both conversational agents. It may also create a greater divide between use of GenAI and teamwork because the task may be more engaging for the humans to discuss amongst themselves. Our study is thus specifically applicable to mostly text and image generation design tasks in pairs. However, we believe some of the results of GenAI perception and workflow processes may apply to design tasks in general, depending on the level of engagement.

\subsubsection{Content analysis and analysis methods}
In our studies, we mainly used qualitative approaches to analyze the data from the interviews and workshop recordings. Other dimensions of data that can provide insight include the content of what is being produced, looking at specific properties like colors and arrangements that may hint at how GenAI and humans effectively designed the outcomes visually. We may also analyze the prompting text written and compare the phrases and word choices that people used with each other during the engagement vs. those used to co-prompt with GenAI. These verbal behavior patterns may provide clues as to how people perceive collaboration vs. co-prompting.

\section{Conclusion}\label{sec:Conclusion}
Our study explored the interplay between co-prompting and collaboration, and how teams perceive GenAI in the team-AI creative collaboration process. To observe the collaboration dynamics, we designed an intevervention based on art design for stage performance and paired students and designers with performance and design backgrounds to work collaboratively with GenAI.

Our findings suggest that the involvement of GenAI in teamwork for creative tasks could be a double-edged sword for team performance. While GenAI can facilitate both the creative process and consensus-building, it can also introduce challenges in communication and strategic prompting. We also found that participants tend to prioritize ideas and suggestions from human collaborators over those from GenAI, suggesting a hierarchy of priority within the team-AI collaboration.

This work illustrates the importance of human-human collaboration when working with GenAI tools. While GenAI can enhance efficiency and provide inspiration, it is most effective when integrated into a collaborative framework that harnesses perspectives and abilities of humans. These findings provide practical guidance to researchers and practitioners involved in developing collaborative design tools in other domains. It highlights the value of understanding human-human collaboration in the context of GenAI.


\bibliographystyle{ACM-Reference-Format}
\bibliography{references}

\end{sloppypar}
\end{document}